\documentclass[a4paper,11pt]{article}
\usepackage[utf8]{inputenc}
\usepackage{anysize}
\marginsize{3cm}{3cm}{2cm}{2cm} 

\usepackage{fancyhdr} 
\usepackage{lipsum} 
\usepackage{xcolor} 
\usepackage{bm} 
\usepackage{amsfonts,amsthm,amsmath,amssymb} 
\usepackage{latexsym,graphicx,epstopdf} 
\usepackage[hyphens]{url}
\usepackage{hyperref}
\usepackage{multirow}
\usepackage[style=numeric,maxbibnames=99,giveninits=true]{biblatex}
\DeclareNameAlias{sortname}{family-given} 
\AtEveryBibitem{\clearfield{month}}
\AtEveryBibitem{\clearfield{day}}

\usepackage[perpage]{footmisc}

\title{The evolution of the COVID-19 pandemic in Chile during 2020: a data perspective
\thanks{This work was partially supported by Hibring Ingenier\'ia.
}}

\author{Gonzalo A. Benavides\thanks{Centro de Investigaci\'on en Ingeniería Matemática, Universidad de Concepci\'on, Concepci\'on, Chile}\textsuperscript{\,\,\,, }{\footnotemark[4]}
\quad Francisca Larach\thanks{Cl\'inica Las Condes, Santiago de Chile, Chile}\\
Vicente Marchant\thanks{Departamento de Ingeniería Matemática, Universidad de Concepción, Concepción, Chile}
\quad Joaquín Fernández\thanks{Departamento de Ingeniería Mecánica, Universidad del Bío-Bío, Concepción, Chile}\textsuperscript{\,\,\,, }{\footnotemark[2]}
\quad Fredy Montoya\thanks{Departamento de Medicina, Universidad de Concepci\'on, Concepci\'on, Chile}\\
Sebasti\'an Dom\'inguez\thanks{Department of Mathematics and Statistics, University of Saskatchewan, Saskatoon, Canada}
\quad Camilo Mejías\thanks{Hibring Ingenier\'ia, Concepci\'on, Chile}\textsuperscript{\,\,\,\,\,\,, }{\thanks{Corresponding author: \href{mailto:cimejias@hibring.cl}{cimejias@hibring.cl}}}
}

\begin{document}

\maketitle

\begin{abstract}
The COVID-19 pandemic has had a great impact in most countries worldwide, in particular impacting Chile greatly to become one of the worst hit countries in the world, despite its low population compared to other affected countries around the world.
In this study we report the evolution of the COVID-19 pandemic and the spread of the SARS-CoV-2 virus in Chile since the first positive case was announced on March 3, 2020, and until November 30, 2020.
We provide a detailed description of the data provided by the Chilean Ministry of Science per administrative region on the number of new cases, tests per capita, and deaths in the country.
\end{abstract}


\vspace{.25cm}


\section{Introduction}
In December 2019, a group of patients from Wuhan, China, developed a respiratory set of symptoms, including atypical pneumonia and respiratory failure.
A novel coronavirus, the SARS-CoV-2, was later found to be the cause of this new disease \cite{Cyranoski2020,Huang2020}. It was named COVID-19 by the World Health Organization (WHO) \cite{WHO2020a}.
A warning on the risk that this disease could create to the global health was sent around the world as early as January 14, 2020 \cite{Hui2020, Bogoch2020}.
The virus rapidly spread to most countries in the world. The WHO declared the COVID-19 disease a global pandemic on March 11, 2020 \cite{WHO2020b}.
Most countries have suffered severely from the COVID-19 pandemic. Spreads of the virus shifted from its original epicenter in China, to Europe, then the US, and later to Latin America.
In mid-July, just over five months after the virus emerged in Wuhan, the Latin America’s death toll from coronavirus surpassed North America’s one, with Peru and Brazil as the epicenters accounting for most of the deaths in the region.

The RT-PCR test (Reverse Transcription Polymerase Chain Reaction) has been the main tool to diagnose COVID-19 active infection.
It is highly specific with no false positive results, but its sensitivity has not been well studied \cite{Corman2020, Tang2020}; it may present false negative results, depending on the stage and severity of the disease, and the quality of the sample \cite{DiPaolo2020,Hernandez-Huerta2020,Woloshin2020}.
False negative results may have serious consequences, as infected people, who might be asymptomatic, may not be isolated, and therefore continue to propagate the infection.

This virus can be transmitted by presymptomatic, symptomatic, or asymptomatic individuals \cite{Rothe2020,Bai2020,Jiang2020}.
Super-spreaders may transmit the virus to hundreds or even thousands of people \cite{Kwok2020}; in some places it has been estimated that 19\% of the cases seeded as much as 80\% of all local transmission \cite{Adam2020}.
The mean incubation time for this disease is 5 days \cite{Bi2020,Nie2020}.
Between 5\% to 20\% of the ill individuals are hospitalized within 14 days from contagion; see, e.g. \cite{Weiss2020}. 

Chile, with a population of approximately 19 million, is divided in 16 administrative regions with a centralized government.
The majority of its inhabitants reside in the Metropolitan, Valparaíso and Biobío regions; these concentrate approximately 60,3\% of the national population (see \cite{ine2017}).
The first COVID-19 case in Chile, a 33-year old man returning from vacations outside Chile, was reported in Talca in the Maule region, on March 3, 2020.
Soon after this first positive case, preventive measures were put in place: on March 14, 2020 in-person classes in schools and universities were suspended, non-essential businesses were mandated to close, and overnight curfews were established along the country.
On March 18, 2020 incoming international flights were restricted to only returning citizens, and domestic traffic decreased greatly.

Until now the Chilean government's strategy to control the epidemic has consisted in testing, tracing, and isolating positive and suspicious cases, the TTA (for its acronym in Spanish) strategy. The government defines positive cases as people whose RT-PCR test result is positive and suspicious cases as those who showed symptoms or were a close contact with a confirmed positive case.
This has also been combined with partial or total lockdowns, applied by municipalities.
Despite all these measures, on the 10th week after the first case, Chile started to suffer a rapid growth in new reported cases, reaching 300,000 cases in the first week of July, overtaking, for example, badly-hit European countries such as Italy and Spain. Soon Chile became one of the most affected countries in the world in terms of confirmed cases and number of deaths; see, e.g., \cite{Shams2020}.
As of November 30, 2020, Chile reached 551,743 infected individuals, and the total number of virus-related fatalities is 15,410, according to official reports.

The objective of this work is to describe the evolution of the epidemic in the Chilean territory during the first nine months, from March 2020 to November 2020, following the first reported case.
We analyze the situation of each region independently and comparatively, in terms of the main statistical parameters used in epidemiology. These include time evolution of new daily cases, testing capacity and fatality rate, providing several and diverse charts that illustrate the data. 

The rest of this paper is organized as follow: in \autoref{section:methods} we describe the methodology used in this study.
In \autoref{section:results} we present our results and provide brief comments on these.
In \autoref{section:discussion} we provide a discussion on the data we present.
We finally arrive at some conclusions in \autoref{section:conclusions}.

\section{Methodology}\label{section:methods}
We utilized Python 3.8 to handle the data and generate all figures presented in this study.

\subsection{Data acquisition}
The Chilean Ministry of Health reports new cases on a daily basis. A new case is defined as a patient who tested positive for COVID-19 with an RT-PCR test.
Since April 29, 2020 Chile began to report both symptomatic and asymptomatic positive cases separately.
All of this information is available at the Github repository from the Chilean Ministry of Sciences \cite{mincienciagithub}.
We consider data corresponding since the first case was reported in the country, until November 30, 2020.
This data does not specify the number of tested individuals.

Regarding to the nature of the data, we point out that both daily new cases and daily RT-PCR tests exhibit oscillatory behaviours.
We ascribe this to the fact that many laboratories that analyze RT-PCR tests in the country do not operate during weekends.
Hence, in order to facilitate the analysis, we smooth out the data by considering a 7-day central moving average.
Both the original and smoothed curves are shown in the corresponding figures presented throughout this article, the former in lighter tones.

\subsection{Data of new cases}
The notification system {\it Epivigila} is a surveillance system utilized by physicians and specialists in Chile. It includes the results for positive cases of all notifiable diseases in Chile.
COVID-19 was considered a disease within this surveillance system by the Chilean law on February 7, 2020 \cite{decreto4}.
Epivigila has received all RT-PCR positive results for COVID-19 in the country every day since then. New COVID-19 cases are only counted when the corresponding RT-PCR test results are positive.
The identification of positive cases is managed with the personal identification number (or R.U.N.~for its acronym in Spanish), preventing duplication.
The statistics on fatalities, on the other hand, include not only RT-PCR confirmed patients, but also those who were suspected to have died from COVID-19 or those who had COVID-19 and died from other pathologies, according to the doctor that filled out the death certificate.

\section{Results}\label{section:results}
\begin{figure}[!ht]
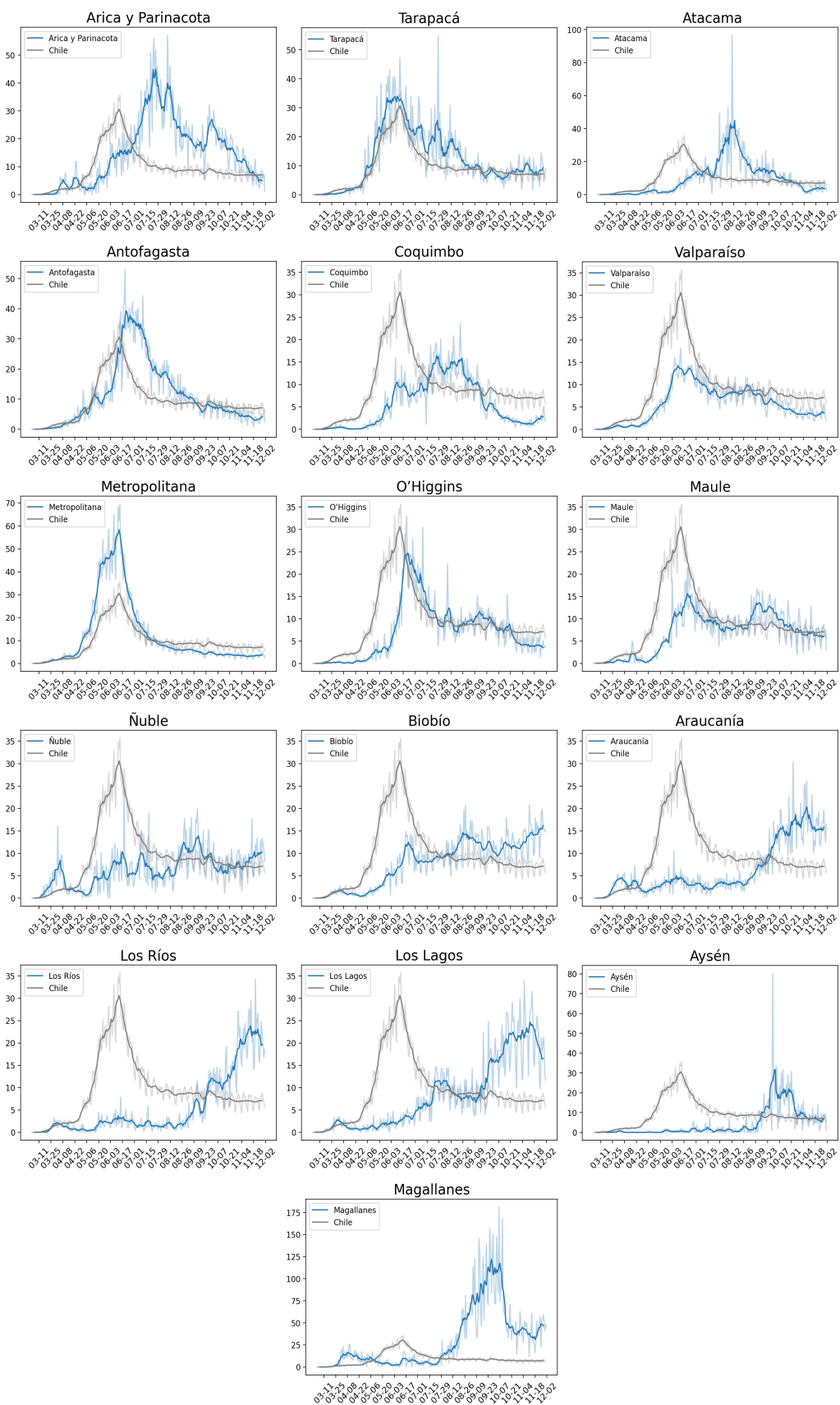

\centering\includegraphics[width = .3\textwidth, height=0.25\textwidth]{AricayParinacota.png}\includegraphics[width = .3\textwidth, height=0.25\textwidth]{Tarapaca.png}\includegraphics[width = .3\textwidth, height=0.25\textwidth]{Atacama.png}
\centering\includegraphics[width = .3\textwidth, height=0.25\textwidth]{Antofagasta.png}\includegraphics[width = .3\textwidth, height=0.25\textwidth]{Coquimbo.png}\includegraphics[width = .3\textwidth, height=0.25\textwidth]{Valparaiso.png}
\centering\includegraphics[width = .3\textwidth, height=0.25\textwidth]{Metropolitana.png}\includegraphics[width = .3\textwidth, height=0.25\textwidth]{OHiggins.png}\includegraphics[width = .3\textwidth, height=0.25\textwidth]{Maule.png}
\centering\includegraphics[width = .3\textwidth, height=0.25\textwidth]{Nuble.png}\includegraphics[width = .3\textwidth, height=0.25\textwidth]{Biobio.png}\includegraphics[width = .3\textwidth, height=0.25\textwidth]{Araucania.png}
\centering\includegraphics[width = .3\textwidth, height=0.25\textwidth]{LosRios.png}\includegraphics[width = .3\textwidth, height=0.25\textwidth]{LosLagos.png}\includegraphics[width = .3\textwidth, height=0.25\textwidth]{Aysen.png}
\centering\includegraphics[width = .3\textwidth, height=0.25\textwidth]{Magallanes.png}
\caption{New cases normalized by 100,000 inhabitants, per region.}\label{fig:NewCasesReg}
\end{figure}

During the following 8 weeks after the first reported case, only few more cases were found in the country.
In \autoref{fig:NewCasesReg} we observe a first small outbreak during March and the first days of April, mainly in the Ñuble, Arica y Parinacota, Araucanía, and Magallanes regions, with an incidence number greater than 5 per 100,000 inhabitants.
Magallanes reached about 20 daily new cases at the beginning of April, then slowly decreased its incidence number during the following two months.
Smaller outbreaks were seen in the Biobío, Los Lagos and Los Ríos regions, with an incidence number of less than 5 per 100,000 inhabitants.
Meanwhile, the rest of the regions of the country experimented no significant rise in the number of positive cases.

\graphicspath{{./Images/NewCasesRegion/}} 

The country’s main wave started within the first days of May.
It affected principally the Metropolitan, Tarapacá, and Antofagasta regions.
During this wave, Chile reached a peak of 35 daily new cases per 100,000 inhabitants at mid-June, the highest recorded up to date.
The Metropolitan region reached about 60 daily new cases per 100,000 inhabitants, whereas Tarapacá's maximum incidence number was roughly 35 daily new cases per 100,000 inhabitants.
On the other hand, the region of Antofagasta culminated during the second half of June in almost 40 daily new cases per 100,000 inhabitants.
By the end of May, the Arica y Parinacota, Valparaíso, Maule, and O’Higgins regions saw the beginning of their first important waves of infection, reaching a maximum incidence number of approximately 15, 15, 15, and 25 daily new cases per 100,000 inhabitants at mid-July, respectively.
The Coquimbo, Biobío, and Ñuble regions took a minor role in the country's main wave, all of them reaching  around 10 daily new cases per 100,000 inhabitants starting June.

During July, the incidence number grew in the Arica y Parinacota and Atacama region, and in a lesser extent in the region of Coquimbo.
They reached about 45, 40, and 15 daily new cases per 100,000 inhabitants at the end of the month. Starting September, new waves were observed in Arica y Parinacota, and Maule regions.
At the same time, at the beginning of this month, a considerable increment of daily new cases took place in the southern regions of Araucanía, Los Ríos, Los Lagos, Aysén, and Magallanes.
As of November 30, 2020 the incidence number in the Araucanía, Los Ríos, and Los Lagos regions appear to have peaked at 15, 20, and 20 daily new cases per 100,000 inhabitants, respectively.  In the Aysén region, on the other hand, the incidence number decreased to under 10 daily new cases per 100,000 inhabitants.
In the region of Magallanes the incidence number surpassed the 100 daily new cases per 100,000 inhabitants---the highest incidence number seen so far in a region of Chile---and appeared to have stabilized at around 40 daily new cases per 100,000 inhabitants starting mid-October.

During the whole period of study, the incidence number in Ñuble has remained low compared to other the regions mentioned above and erratically between 5 and 15 daily new cases per 100,000 inhabitants since mid-May, with no clear waves. The Biobío region, on the other hand, also presents a special behaviour, with an incidence number increasing slowly and consistently from 10 daily new cases per 100,000 inhabitants at the end of June, to 15 by the end of November.

\begin{figure}[!ht]
\centering\includegraphics[width = .75\textwidth, height=.5\textwidth]{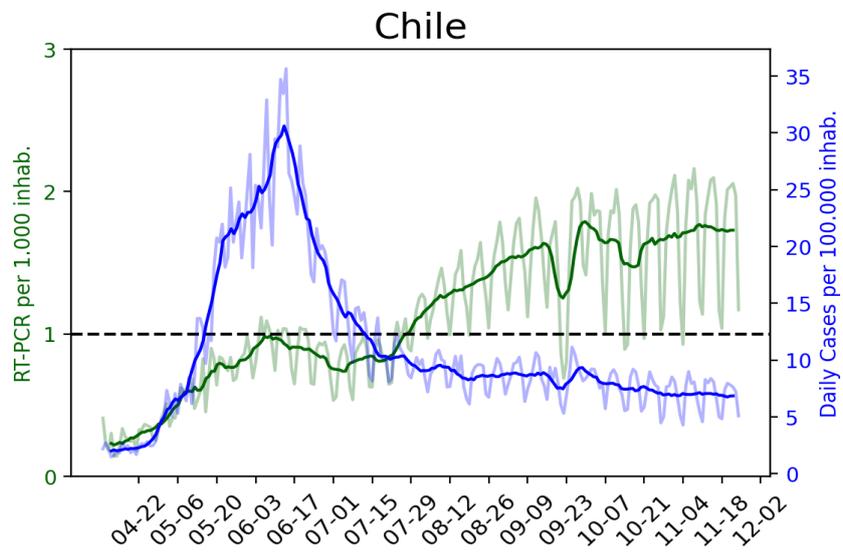}
\caption{RT-PCR and daily cases per capita.}\label{fig:RT-PCRChile}
\end{figure}

\begin{figure}[!ht]
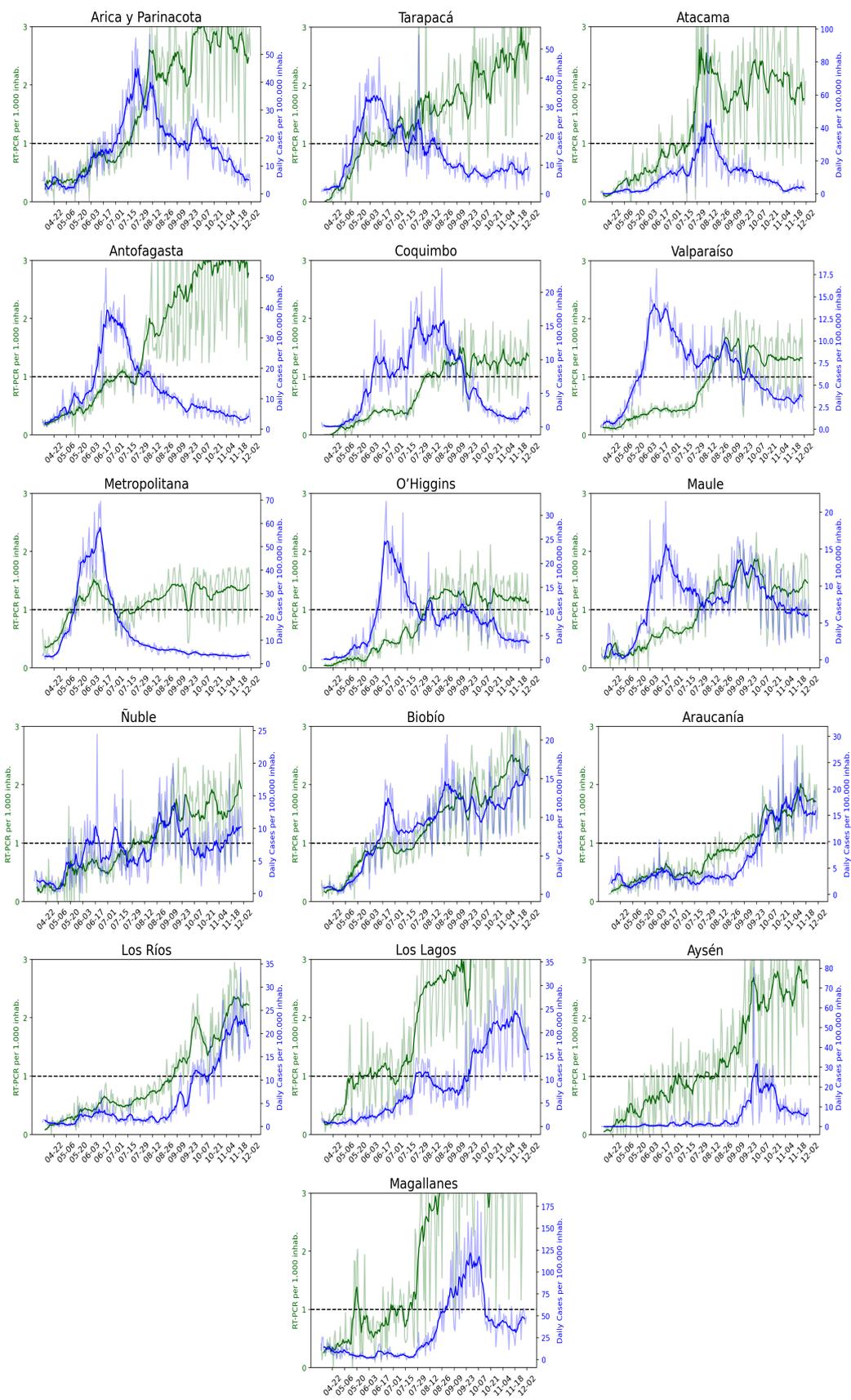

\centering\includegraphics[width = .3\textwidth, height=0.25\textwidth]{AricayParinacota.png}\includegraphics[width = .3\textwidth, height=0.25\textwidth]{Tarapaca.png}\includegraphics[width = .3\textwidth, height=0.25\textwidth]{Atacama.png}
\centering\includegraphics[width = .3\textwidth, height=0.25\textwidth]{Antofagasta.png}\includegraphics[width = .3\textwidth, height=0.25\textwidth]{Coquimbo.png}\includegraphics[width = .3\textwidth, height=0.25\textwidth]{Valparaiso.png}
\centering\includegraphics[width = .3\textwidth, height=0.25\textwidth]{Metropolitana.png}\includegraphics[width = .3\textwidth, height=0.25\textwidth]{OHiggins.png}\includegraphics[width = .3\textwidth, height=0.25\textwidth]{Maule.png}
\centering\includegraphics[width = .3\textwidth, height=0.25\textwidth]{Nuble.png}
\includegraphics[width = .3\textwidth, height=0.25\textwidth]{Biobio.png}
\includegraphics[width = .3\textwidth, height=0.25\textwidth]{Araucania.png}
\centering\includegraphics[width = .3\textwidth, height=0.25\textwidth]{LosRios.png}\includegraphics[width = .3\textwidth, height=0.25\textwidth]{LosLagos.png}\includegraphics[width = .3\textwidth, height=0.25\textwidth]{Aysen.png}
\centering\includegraphics[width = .3\textwidth, height=0.25\textwidth]{Magallanes.png}
\caption{Evolution of RT-PCR testing by region.}\label{fig:NewCasesRT-PCR}
\end{figure}

\autoref{fig:RT-PCRChile} and \autoref{fig:NewCasesRT-PCR} show the evolution of both the testing capacity and daily new cases in Chile as a whole, and for each of the 16 major administrative regions in the country.
In \autoref{fig:RT-PCRChile} we observe that the testing capacity of the Chilean healthcare system remained under 0.5 daily RT-PCR tests per 1,000 inhabitants until the end of March. At the beginning of April, it started to increase steadily, and reached approximately 0.8 in the first days of June during the peak of new daily new cases in Chile (cf.~\autoref{fig:NewCasesReg}).
The majority of these tests were made in the Metropolitan Region (cf.~\autoref{fig:NewCasesRT-PCR}).
After remaining almost constant for about a month, daily RT-PCR started to increase rapidly again up to roughly 1.5 in mid-August.
The testing capacity continued to increase slowly to reach a stable value of approximately 1.7 daily tests per 1,000 inhabitants in mid-October.
There is a remarkable exception in the increase of number of tests performed daily during the national holidays, celebrated September 18 to September 20, 2020 where the testing capacity decreased momentarily. 
The northern regions Arica y Parinacota, Tarapacá, Atacama, and Antofagasta, and the southern regions, Los Lagos, and Magallanes, showed a rapid increase in testing capacity during July, surpassing 1.5 daily RT-PCR tests per 1000 inhabitants at the end of the month.
Some of these regions, Arica y Parinacota, Antofagasta, Los Lagos and Magallanes were even able to keep over 2 daily RT-PCR per 1000 inhabitants until the end of the period of study.
Of the latter, the Magallanes region has attained the highest testing capacity among all the 16 regions of Chile, surpassing 3 daily RT-PCR per 1000 inhabitants since late August.
The regions Los Ríos and Aysén experimented later rises in testing capacity starting from the end of August, they attained and maintained values over 1.5 daily RT-PCR tests per 1000 inhabitants since the beginning of October.
On the other hand, the central regions Coquimbo, Valparaíso, Metropolitan, O'Higgins and Maule have also experimented an increment in their testing capacity, but in a more conservative way reaching approximately 1.3 daily RT-PCR tests per 1000 inhabitants.
Finally, the Biobío and Ñuble regions have shown a sustained increment in their testing capacity since the beginning of the epidemic in Chile, attaining approximately 2 daily RT-PCR tests per 1000 inhabitants, respectively, at the beginning and end of November.

The total fatalities and Case Fatality Ratio (CFR) per region as of November 30, are presented in \autoref{table:coviddeaths}.
The Metropolitan region concentrates the majority of the fatalities with approximately 67\% of the total deaths related to COVID-19.
The Valparaíso and Metropolitan regions have the highest CFR, with 3.5 and 3.4, respectively.

\begin{table}
    \centering
    \begin{tabular}{|c|c|c|c|c|}
        \hline
        \multirow{2}{*}{Region} & Population & Total Cases & Total Deaths & \multirow{2}{*}{CFR} 
        \\ & & as of 11-03 & as of 11-03 &
        \\\hline
        Arica y Parinacota & 252,110 & 10,239 & 207 & 2.0 \\\hline
        Tarapacá & 382,773 & 13,876 & 258 & 1.9\\\hline
        Antofagasta & 691,854 & 22,226 & 562 & 2.5\\\hline
        Atacama & 314,709 & 8,161 & 106 & 1.3 \\\hline
        Coquimbo & 836,096 & 13,132 & 270 & 2.1 \\\hline
        Valparaíso & 1,960,170 & 32,441 & 1151 & 3.5 \\\hline
        Metropolitana & 8,125,072 & 305,218 & 10,342 & 3.4\\\hline
        O'Higgins & 991,063 & 19,857 & 544 & 2.7\\\hline
        Maule & 1,131,939 & 21,095 & 489 & 2.3 \\\hline
        Ñuble & 511,551 & 8,812 & 178 & 2.0\\\hline
        Biobío & 1,663,696 & 36,599 & 607 & 1.7\\\hline
        Araucanía & 1,014,343 & 17,531 & 229 & 1.3\\\hline
        Los Ríos & 405,835 & 6,105 & 55 & 0.9\\\hline
        Los Lagos & 891,440 & 20,667 & 203 & 1.0\\\hline
        Aysén & 107,297 & 1,269 & 14 & 1.1\\\hline
        Magallanes & 178,362 & 14,470 & 194 & 1.3\\\hline
        \hline
        Total & 19,458,310 & 551,698 & 15,409 & 2.8 \\\hline
    \end{tabular}
    \caption{Total fatalities and CFR per region as of November 30.}
    \label{table:coviddeaths}
\end{table}

\graphicspath{{./Images/NewCases_RT-PCR/}} 

\graphicspath{{./Images/ContagiosPCRFinal/}} 

\graphicspath{{./Images/}} 
\begin{figure}[!ht]
\centering\includegraphics[width = 0.9\textwidth, height=.6\textwidth]{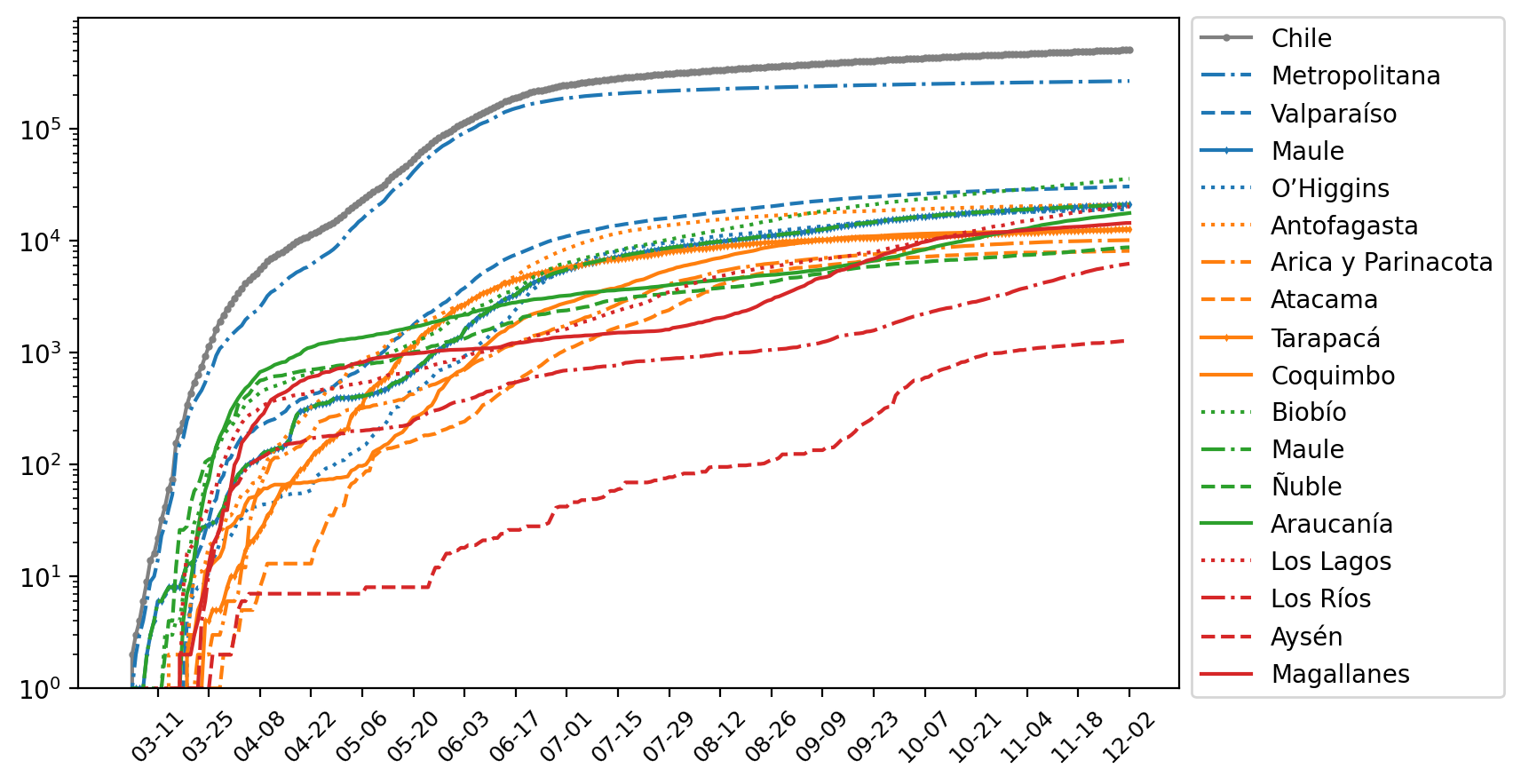}
\caption{Total number of confirmed cases by region (in semilog scale).}\label{fig:ClustersNewCases}
\end{figure}

\section{Discussion}\label{section:discussion}
Chile has been one of the most affected countries in the world by the COVID-19 disease, in terms of number of cases and fatalities per capita.
By July, 2020 Chile had the highest number of total cases per capita in the world.
It was also the ninth country in the world in terms of fatalities per capita; \cite{Shams2020}.
In addition, Chile became the sixth country in the world by means of number of confirmed cases \cite{ChileSexto}.
One of the reasons for this phenomenon could be the weather, since the first wave of COVID-19 arrived to Chile  at beginning of the autumn season, similarly to what it was recently reported in Norway; see, e.g.~\cite{Menebo2020}.
From \autoref{section:results} it transpires that the epidemic in Chile can be described as asynchronous and diverse in nature in each of its 16 major administrative regions.
These differences may be explained by several factors.
Some of them are the great differences in geography, population density, and urbanization of each region.
While the northernmost regions are mainly dominated by arid weather, the central regions enjoy Mediterranean weather.
The rest of the country, on the other hand, is ruled by an oceanic and colder weather.
Moreover, some regions are highly urbanized (e.g.~Metropolitan region), whereas others still have a great amount of people living in rural areas (e.g.~Ñuble region); see, for instance, \cite{ine2017}.
Differences in transport connectivity between regions may have also played a major role in the evolution of the epidemic.
The Chilean territory from the north to the Los Lagos region is well connected by land, whereas Aysén and Magallanes regions are virtually isolated in this sense.
With respect to air traffic, the major airports of the country located in the Arica y Parinacota, Metropolitan, Biobío and Tarapacá regions.
They have served as important focuses of infection at the beginning of the spread; the restrictions in international and domestic flights were only applied after 11 days after the report of the first confirmed case.
We highlight that the Metropolitan region's main airport as the main source of imported contagions, as it is the principal international airport in Chile.
Regardless of these restrictions, many routine flights, considered essential for the nation, continued operating during the whole period of study.

The first contagions of the country were most probably of imported origin.
In fact, the first confirmed case of COVID-19 in Chile was an individual returning from vacations in Southeast Asia.
The majority of the subsequent cases along the country during March and April, 2020 corresponded to returning citizens, foreign tourists, and people who interacted with them.
In mid-March, an outbreak was declared at a gymnasium in the city of Chillán, in the Ñuble region \cite{Chillangimnasio}; 21 people were confirmed to have been infected by a foreign visitor.
Another 6 people were later linked to this same outbreak.
By the end of March and the beginning of April, the Magallanes region suffered its first wave.
The Chilean government soon conjectured that the outbreak was due to the arrival of several cruise ships transporting tourists of diverse nationalities \cite{Magallanescruceros,Magallanescruceros2}.
After the report of 155 confirmed cases in the country, the Chilean government acknowledged {\it community transmission} of the SARS-CoV-2 virus along the country, and announced the closure of the Chilean borders in March 16, 2020 \cite{MinsalMarzo16}.

In mid-May, about two weeks after the beginning of the main wave in the Metropolitan region, the neighboring regions of Coquimbo, Valparaiso, O'Higgins and Maule also initiated their main waves.
Much probably, they got heavily contaminated by the Metropolitan region due to land proximity and easy land connectivity, as they do not have important airports.
The Coquimbo region, interestingly, was able to initially contain this outbreak during mid-June, which later resurfaced in July and August.

Simultaneously to the aforementioned regions, the northern regions of Tarapacá and Antofagasta, distant from the Metropolitan Region, also suffered rapid outbreaks.
This behavior is probably due to maintained air traffic with the Metropolitan region for mining activity issues, which did not stop for being considered essential.
The Atacama region, also with important mining activity, was able to initially avoid rapid increments in daily new cases.
Although, during July and August, a violent increment in new cases was seen.

On the other hand, the northernmost region Arica y Parinacota also started to suffer a rapid increase in its daily new cases at the end of May.
This outbreak was controlled during June, to then heavily reappeared during July and August.
Despite not having important mining centers, the region possesses one of the fourth most important airports in the country, strongly connected with the one of the Metropolitan region and the virtually only way of passenger transport to central Chile.
Most surely, the virus reached the region through this airport.

The southern regions of Los Ríos and Araucanía managed to keep the incidence below 5 daily new cases per 1000 inhabitants until mid-August.
The Los Lagos region accomplished similar results until the beginning of July.
These behaviors are fairly impressive as winters there are usually rough.
We assume this slow spread was due to better isolation and preparation, thanks to the experience gained from the central and northern Chile.

The Aysén region, the least-populated and second vastest region of the country, had reported only 124 cases in total (about 0.1\% of its population) until the beginning of September, when a rapid outbreak started.
We consider that the region managed to practically seclude from the virus for a long period of time due to its virtual land-isolation and low rates of stranger arrivals.
It is important to notice that, in pre-pandemic times, the region characterizes for receiving a great amount of foreign tourists every year.

The Ñuble region, has evidenced the most erratic behaviour among all the 16 regions of Chile, with constant increases and decreases in the number of daily new cases.
We explain this phenomenon based on the fact that approximately 30.6 \% of the population in the Ñuble region lives in rural areas, making it the most rural region of the country \cite{ine2017}.
We speculate that great part of small waves presented during the period of study correspond to localized outbreaks, which were effectively contained mainly due to the facility in isolation that rurality provides.
The fact that there is no airport in the region may have also served to prevent further spread.

With about 9\% of its population already infected and an incidence above 110 daily new cases per 1000 inhabitants during mid-September, the Magallanes region has been the most affected region of the country.
The majority of the epidemic in the region is explained by its capital Punta Arenas, which concentrates approximately 75\% of the population of the Magallanes region.
We speculate that, after the recovery from the outbreak experimented in June-July---which was part of the national's main wave---and due to the consequent resumption of great part of flights connecting with the rest of the regions (especially the Metropolitan region), Magallanes received a considerable number of imported cases.
This, combined with the fact that the principal trading and administrative activity of the region are concentrated in a few blocks of Punta Arenas, might have triggered the enormous outbreak.

Chile has done an outstanding effort in enhancing the current molecular biology laboratories and opening new public, private, university-based laboratories.
The country has shown a sustained increase in testing capacity during the whole period of study.
As of October 23, 2020, this has allowed Chile to have the first place in testing per capita capacity in Latin America \cite{ChilePCRprimero}, keeping over 1 daily RT-PCR per 1,000 inhabitants 
since the beginning of August and stabilizing in about 1.8 daily RT-PCR per 1,000 inhabitants during November (see \autoref{fig:RT-PCRChile}).
At the same time, the utilization of the unified patient database notification system Epivigila, have made it possible to ensure a faithful and strong database.

Nevertheless, we stress that much probably there was a significant subreport during the the beginning of the pandemic, especially during the main wave that took place in May-June; the testing capacity there was below 1 daily RT-PCR per 1,000 inhabitants.
This initial deficit in testing capacity, due to lack of proper instalments and equipment, resulted in that testing was performed predominantly in symptomatic cases, leaving virtually apart close contacts and with scarce active search strategies.

During the peak period of the first wave (mid May to end of June) the rapid growth in testing capacity caused delays in rendering times due to both delay in the processing of RT-PCR tests and the subsequent rendering of results to available centralized databases.
For this reason, in this study we do not consider active cases for our reports due to the differences in waiting time for RT-PCR results, with delays that could result in many new cases counted as such only after their 14 days isolation period had ended.

The results of containment strategies have varied significantly among the country.
Most of the regions required lockdowns at a municipality level, with the exception of the Metropolitan Region that required an early complete lockdown for 2 months, and was even longer for certain municipalities within it.
Highest population density and dense transportation system could explain the biggest outbreak in Metropolitan Region and the difficulty of containing/restraining it.
We analyze outbreak dynamics at a regional level, so we cannot see inter municipality differences.
According to \cite{Bennett2021}, different socioeconomic conditions and mobility can play an important role in differences in lockdown effectiveness that are seen when comparing municipalities from one same region.
Despite the country's preparedness, and the good development status of ICUs in Chile, a high CFR was observed as of November 30, 2020 (see \autoref{table:coviddeaths})---the CFR of the A/H1N1 influenza epidemic in Chile in 2009 is estimated in 0.9 \% \cite{influenzaChile2009}.
The cause for this is probably multifactorial: High incidence of comorbidities such as overweight and obesity (about 74 \% of the adult population are overweight or obese \cite{obesidadChile}), hypertension (approximately 27.6 \% of the population suspects of suffering hypertension \cite{EncuestaSaludChile}), and smoking (33.3 \% of the population smoke regularly \cite{EncuestaSaludChile}), among others;
health system saturation during peak period, especially emergency rooms, and delay in consultation of severe cases due to population's ignorance about the severity of the sickness.
In \cite{Chaudhry2020} it was found that increasing COVID-19 caseloads were associated with countries with higher obesity, higher median population age and longer time to border closures from the first reported case and increased mortality per million was significantly associated with higher obesity prevalence.
On the contrary, per capita gross
domestic product and reduced income dispersion reduced mortality.
Rapid border closures, full lockdowns, and wide-spread testing were not associated with COVID-19 mortality per million people.

\section{Conclusions}\label{section:conclusions}
We presented notable differences in behaviour of outbreak dynamics---incidence, testing capacity and fatalities---among all 16 major administrative regions of Chile.
We argued on geographical and demographics differences as possible explanations to this behavior. We decided to leave out the data corresponding to December 2020 as the holidays impacted the behaviour of the pandemic in a great manner due to the holidays (including shopping, gatherings, travelling, and more).

\paragraph{Conflict of interest}
The authors declare no conflicts of interest.



\printbibliography

\end{document}